# Waveguide optical parametric amplifiers in silicon nitride with 2D graphene oxide films

*Yang Qu[1,#], Jiayang Wu[1, #, *], Yuning Zhang[1], Yunyi Yang[1], Linnan Jia[1], Houssein El Dirani[2, [+]], Sébastien Kerdiles[2], Corrado Sciancalepore[2, [++]], Pierre Demongodin[3], Christian Grillet[3], Christelle Monat[3], Baohua Jia[4, 5 *], and David J. Moss[1, *]*

[1]Optical Sciences Centre, Swinburne University of Technology, Hawthorn, VIC 3122, Australia

[2]Université Grenoble-Alpes, CEA-LETI, 17 Avenue des Martyrs, 38054 Grenoble, France

[+]Current address: LIGENTEC SA, 224 Bd John Kennedy, 91100 Corbeil-Essonnes, France

[++]Current address: Soitec SA, 38190 Bernin, France

[3]Université de Lyon, Ecole Centrale de Lyon, INSA Lyon, Université Claude Bernard Lyon 1, CPE Lyon, CNRS, INL, UMR5270, 69130 Ecully, France

[4]Centre for Atomaterials and Nanomanufacturing, School of Science, RMIT University, Melbourne, VIC 3000, Australia

[5]Australian Research Council (ARC) Industrial Transformation Training, Centre in Surface Engineering for Advanced Materials (SEAM), RMIT University, Melbourne, Victoria 3000, Australia

[#]These authors contribute equally.

*E-mail: *jiayangwu@swin.edu.au*, *baohua.jia@rmit.edu.au*, *dmoss@swin.edu.au*

# Abstract

Optical parametric amplification (OPA) represents a powerful solution to achieve broadband amplification in wavelength ranges beyond the scope of conventional gain media, for generating high-power optical pulses, optical microcombs, entangled photon pairs and a wide range of other applications. Here, we demonstrate optical parametric amplifiers based on silicon nitride ($Si_3N_4$) waveguides integrated with two-dimensional (2D) layered graphene oxide (GO) films. We achieve precise control over the thickness, length, and position of the GO films using a transfer-free, layer-by-layer coating method combined with accurate window opening in the chip cladding using photolithography. Detailed OPA measurements with a pulsed pump for the fabricated devices with different GO film thicknesses and lengths show a maximum parametric gain of ~24.0 dB, representing a ~12.2 dB improvement relative to the device without GO. We perform a theoretical analysis of the device performance, achieving good agreement with experiment and showing that there is substantial room for further improvement. This work represents the first demonstration of integrating 2D materials on chips to enhance the OPA performance, providing a new way of achieving high performance photonic integrated OPA by incorporating 2D materials.

# Introduction

Optical amplifiers are key to many applications[1-3] such as optical communications where they have been instrumental with rare-earth-doped fibers[4-6] and III-V semiconductors[7-9]. However, these devices are restricted to specific wavelength ranges determined by the intrinsic bandgap of the materials[1,10]. In contrast, optical parametric amplification (OPA) can achieve gain across virtually any wavelength range[11,12], and so is capable of achieving broadband optical amplification outside of conventional wavelength windows restricted by intrinsic materials properties[11,13]. Since its discovery in 1965[14], OPA has found applications in many fields such as ultrafast spectroscopy[15,16], optical communications[5,13], optical imaging[17,18], laser processing[19,20], and quantum optics[21,22]. Notably, it has underpinned many new technological breakthroughs such as optical microcombs[23,24] and entangled photon pairs[25,26].

To achieve OPA, materials with a high optical nonlinearity are needed – either second- ($\chi^{(2)}$) or third-order ($\chi^{(3)}$) nonlinearities[27,28], and OPA has been demonstrated in birefringent crystals [29-31], optical fibers[10,32,33], and photonic integrated chips[1,3,24,34,35]. Amongst these, photonic integrated chips offer the advantages of a compact footprint, low power consumption, high stability and scalability, as well as cost reduction through large-scale manufacturing[36-38]. Despite silicon's dominance as a platform for linear photonic integrated devices[39,40], its significant two photon absorption (TPA) in the near infrared wavelength region and the resulting free carrier absorption lead to a high nonlinear loss[3,27], making it challenging to achieve any significant OPA gain in this wavelength range. Other nonlinear integrated material platforms, such as silicon nitride ($Si_3N_4$)[1,41], silicon rich nitride[42,43], doped silica[36,44], AlGaAs[45,46], chalcogenide[47,48], GaP[49], and tantala[50], exhibit much lower TPA at near infrared wavelengths and have made significant progress over the past decade. However, the low third-order optical nonlinearity of some of these platforms such as silicon nitride and doped silica imposes a significant limitation on the OPA gain that they can achieve.

Recently, two-dimensional (2D) materials with ultrahigh optical nonlinearities and broadband response have been integrated on photonic chips to achieve exceptional nonlinear optical performance[25,51-54], highlighted by the progress in realizing OPA by exploiting the high second-order optical nonlinearities of monolayer transition metal dichalcogenides (TMDCs)[25]. Previously[55-59], we reported an ultra-high third-order optical nonlinearity in 2D graphene oxide (GO) films that is about 4 orders of magnitude larger than silicon, together with a large bandgap (> 2 eV) that yields a linear loss more than 2 orders of magnitude lower than graphene, and perhaps most importantly, low TPA at near infrared wavelengths – all of which are key to achieving high OPA. In addition, GO has demonstrated high compatibility with various integrated platforms [12,38], along with the capability to achieve precise control over its film thickness and length[56,60].

In this work, for the first time, on-chip integration of 2D materials to achieve improved OPA performance is demonstrated, which is realized through the integration of 2D layered GO films onto $Si_3N_4$ waveguides. We employ a transfer-free, layer-by-layer coating method to achieve precise control over the GO film thickness, and by using photolithography to open windows in the waveguide cladding we are able to accurately control the GO film length and position. The processes for GO synthesis and on-chip integration are discussed in more detail elsewhere[12,19,54]. We perform a detailed experimental characterization of the OPA performance of the devices with different GO film thicknesses and lengths, achieving a maximum parametric gain of ~24.0 dB, representing a ~12.2 dB improvement over the uncoated device. By fitting experimental results with theory, we analyse the influence of the applied power, wavelength detuning, and GO film thickness and length on the OPA performance, and in the process demonstrate that there is still significant potential for further improving the performance. These results verify the effectiveness of the on-chip integration of 2D GO films to improve the OPA performance of photonic integrated devices.

**Experimental results**

**GO properties.** **Figure 1(a)** illustrates the atomic structure and bandgap of GO, which is a derivative of graphene. Unlike graphene, which consists solely of $sp^2$-hybridized carbon atoms, GO contains various oxygen-containing functional groups (OCFGs) such as hydroxyl, carboxyl, and carbonyl groups[12]. Some of the carbon atoms in GO are $sp^3$-hybridized through σ-bonding with the OCFGs, resulting in a heterogeneous structure. In contrast to graphene, which has a zero bandgap, GO has an opened bandgap resulting from the isolated $sp^2$ domains within the $sp^3$ C–O matrix. The bandgap of GO typically falls between 2.1 eV and 3.6 eV[38], resulting in both low linear light absorption and low nonlinear TPA at near-infrared wavelengths[54]. These properties, along with the high optical nonlinearity of GO[56,58], are essential for nonlinear optical applications such as OPA, which are also the key motivation for the choice of GO for our study. We note that although many other 2D materials have demonstrated significant optical nonlinearity[51,61], the substantial loss caused by their strong light absorption poses a challenge in the pursuit of high net parametric gain. Moreover, the material properties of GO can be tuned by manipulating the OCFGs to engineer its bandgap, which has enabled a range of photonic, electronic, and optoelectronic applications[12].

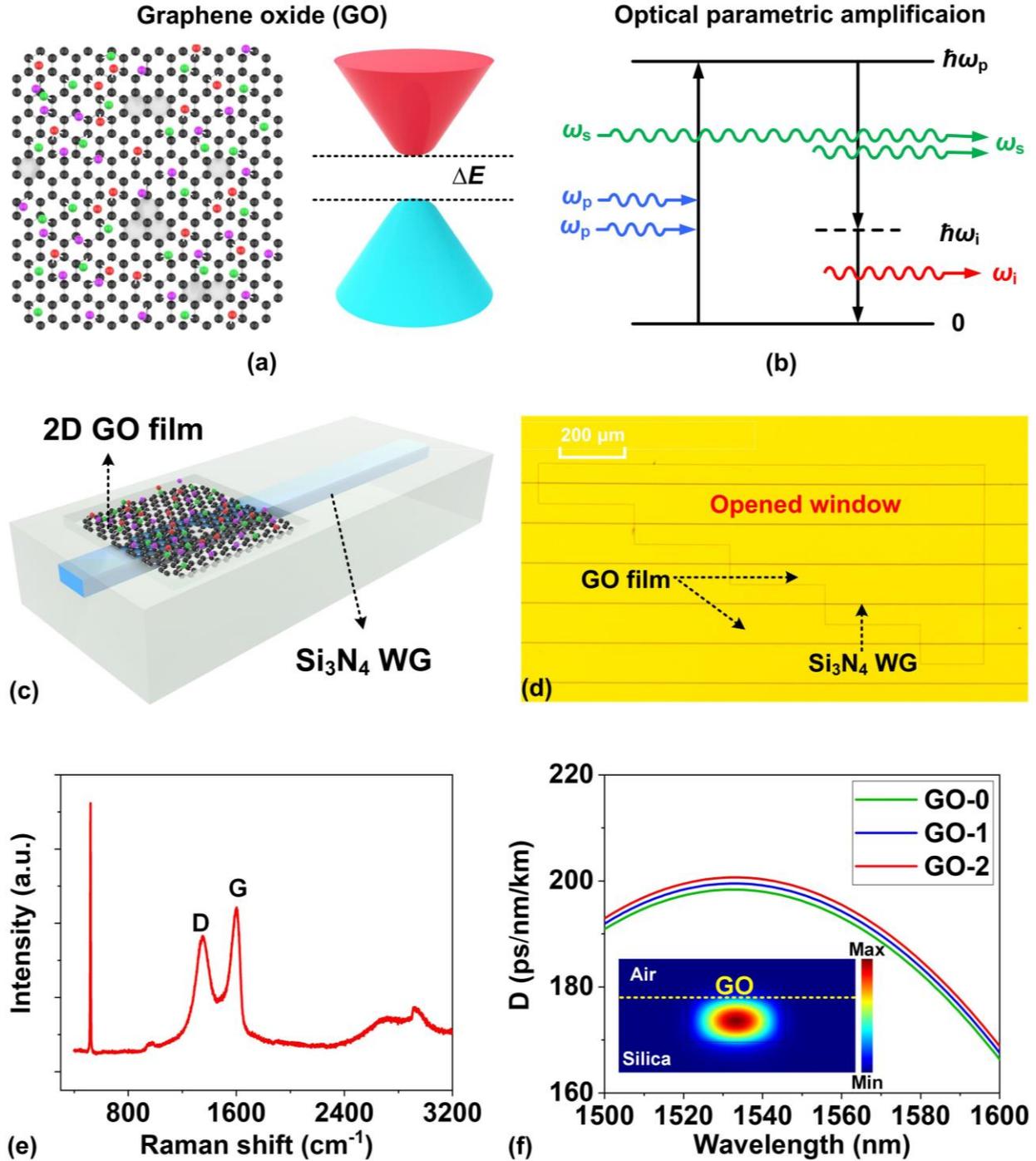

**Figure 1**. (a) Schematic of GO's atomic structure and bandgap. The colorful balls in the atomic structure represent the diverse oxygen-containing functional groups (OCFGs). (b) Schematic of signal amplification based on optical parametric process. (c) Schematic of a $Si_3N_4$ waveguide integrated with a single layer GO film. (d) Microscopic image of the fabricated $Si_3N_4$ integrated chip coated with a single layer GO film. (e) Measured Raman spectrum of the GO-coated $Si_3N_4$ chip in (d). (f) Dispersion ($D$) of the uncoated waveguide (GO-0) and hybrid waveguides with 1 and 2 layers of GO (GO-1, GO-2). Inset shows TE mode profile of the $Si_3N_4$ waveguide integrated with a single layer GO film.

**Figure 1(b)** illustrates the principle of signal amplification based on an optical parametric process[62]. In this process, when pump and idler photons travel collinearly through a nonlinear optical medium, a pump photon excites a virtual energy level. The decay of this energy level is stimulated by a signal photon, resulting in the emission of an identical second signal photon and an idler photon, while conserving both energy and momentum. In processes that involve optical absorption, such as photoluminescence and TPA, real photogenerated carriers are involved, which can alter the quiescent material nonlinear response[12,27]. In contrast, the optical parametric process operates by virtual excitation of carriers, thus avoiding the creation of photogenerated carriers. This makes it quasi-instantaneous, with ultrafast response times on the order of femtoseconds[1,54]. We note that although the parametric gain itself is almost instantaneous, when influenced by nonlinear absorption with much slower recovery times such as that induced by free carriers in silicon[27], the net parametric gain can accordingly have a slow time response component.

**Device design and fabrication. Figure 1(c)** illustrates the schematic of a $Si_3N_4$ waveguide integrated with a single layer GO film. Compared to silicon that has a small (indirect) bandgap of ~1.1 eV[27], $Si_3N_4$ has a large bandgap of ~5.0 eV[36] that yields low TPA in the near-infrared region. To enable the interaction between the GO film and the evanescent field of the waveguide mode, a portion of the silica upper cladding was removed to allow for the GO film to be coated on the top surface of the $Si_3N_4$ waveguide. **Figure 1(d)** shows a microscopic image of the fabricated $Si_3N_4$ chip integrated with a single layer GO film. The successful coating of the GO film is confirmed by the presence of the representative D (1345 $cm^{-1}$) and G (1590 $cm^{-1}$) peaks in the measured Raman spectrum, as shown in **Figure 1(e)**. First, we fabricated low-loss $Si_3N_4$ waveguides via CMOS-compatible processes (see Methods). Next, we coated the waveguides with 2D GO films using a transfer-free, solution-based coating method (see Methods). This approach allows for large-area, layer-by-layer film coating with high repeatability and compatibility with various integrated material platforms[12,38,63]. The thickness of the GO film, characterized via atomic force microscopy measurements, was ~2 nm. The high transmittance and excellent morphology of the fabricated device demonstrate that our GO coating method, based on self-assembly via electrostatic attachment, can achieve uniform film coating in the window opening area without any noticeable wrinkling or stretching. It is also worth noting that our GO film coating method enables conformal film coating, which allows direct contact and enclosing of GO films with integrated waveguides. This yields efficient light–matter interaction and offers advantages compared to film transfer techniques commonly used for coating other 2D materials like graphene and TMDCs[19]. Varying the film thickness also effectively serves to vary the mode overlap with the film without introducing a cladding separation layer. A detailed characterization for the conformally coated GO films is provided in Ref.[64]. The length and position of the GO films can also be easily controlled by adjusting the length and position of the windows opened on the silica upper cladding, which provides high flexibility for optimizing the performance of the hybrid waveguides by altering the GO film parameters.

**Figure 1(f)** shows the dispersion $D$ of the uncoated waveguide and the hybrid waveguides with 1 and 2 layers of GO, calculated with commercial mode solving software using the materials' refractive indices measured by spectral ellipsometry. The $Si_3N_4$ waveguides in all these devices had a cross section of 1.60 μm × 0.72 μm, and the inset in **Figure 1(f)** depicts the transverse electric (TE) mode profile of the hybrid waveguide with 1 layer of GO. The interaction between the waveguide's evanescent field and the highly nonlinear GO film enhances the nonlinear optical response of the hybrid waveguide, which forms the basis for improving the OPA performance. Note that we have discussed the physics and mechanisms for the enhancement in the nonlinear optical performance of GO-coated integrated photonic devices elsewhere as well[54,56,57]. The remarkably high Kerr nonlinearity observed for 2D GO films (with a Kerr coefficient $n_2$ that is about five orders of magnitude higher than $Si_3N_4$[38,55]) originates from the significant heterogeneity within the atomic structure and the strong anisotropy exhibited by the 2D films[12,54].

For the hybrid waveguides with 1 and 2 layers of GO, the GO mode overlaps are ~0.07 % and ~0.15 %, respectively. We selected TE-polarization for our subsequent measurements since it supports in-plane interaction between the waveguide's evanescent field and the GO film, which is much stronger than the out-of-plane interaction due to the significant optical anisotropy in 2D materials[65,66]. In **Figure 1(f)**, it can be observed that all three waveguides exhibit anomalous dispersion, which is crucial for reducing phase mismatch and improving the parametric gain in the optical parametric process. Upon incorporating 1 layer of GO, the hybrid waveguide shows a slightly increased anomalous dispersion compared to waveguides without GO. For the hybrid waveguides with 2 layers of GO, the anomalous dispersion is further enhanced.

**Loss measurements.** The coating of GO films onto $Si_3N_4$ waveguides introduces extra linear and nonlinear loss. Before the OPA measurements, we used the experimental setup in **Figure S1** of the Supplementary Information to characterize the linear and nonlinear loss of the fabricated devices. Fiber-to-chip coupling was achieved via lensed fibers butt coupled to inverse-taper couplers at both ends of the $Si_3N_4$ waveguides. The coupling loss was ~4.2 dB / facet. We measured three devices, including the uncoated $Si_3N_4$ waveguide and hybrid waveguides with 1 and 2 layers of GO. The $Si_3N_4$ waveguides in these devices were all ~20 mm in length, while for the hybrid waveguides, windows with a length of ~1.4 mm were opened at a distance of ~0.7 mm from the input port. In our following discussion, the input light power quoted refers to the power coupled into the devices, with the fiber-to-chip coupling loss being excluded.

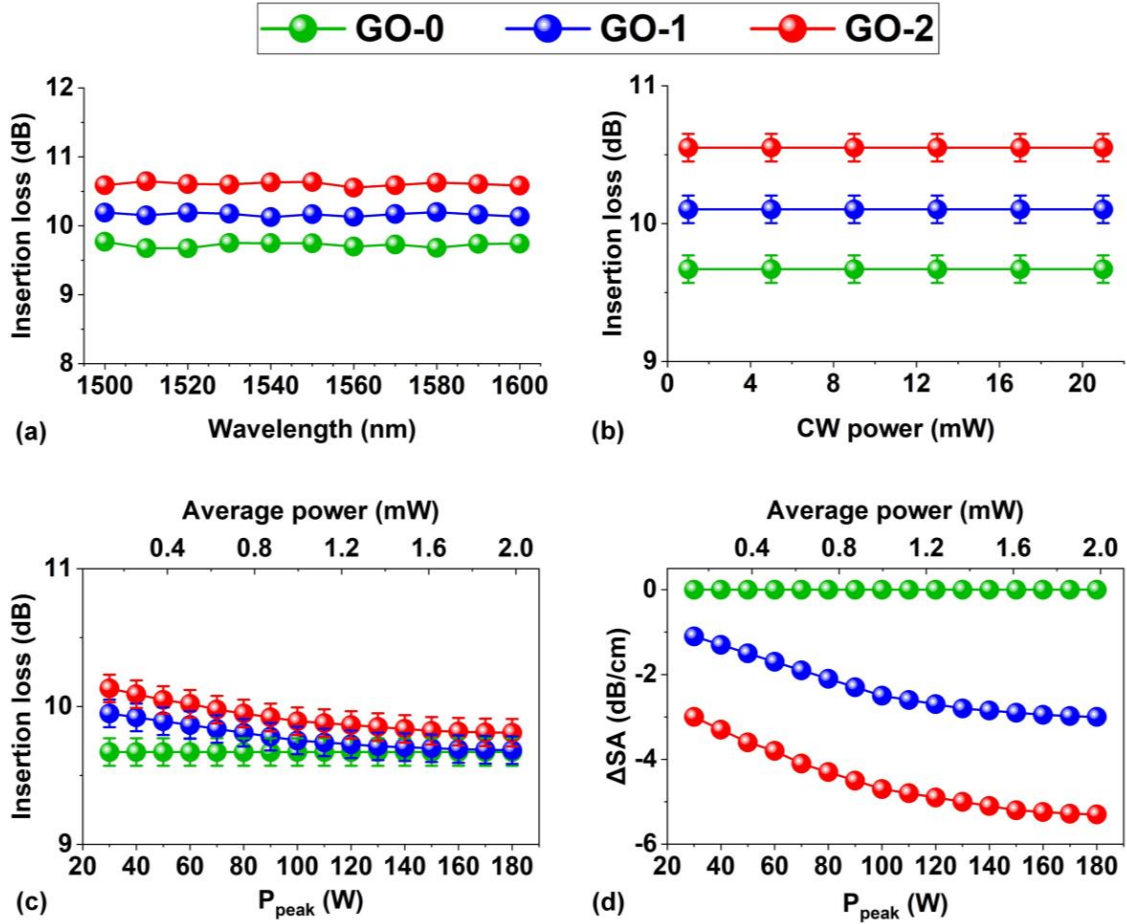

**Figure 2.** Experimental results for loss measurements. (a) Measured insertion loss versus wavelength of input continuous-wave (CW) light. The input CW power is ~1 mW. (b) Measured insertion loss versus input CW power. The input CW wavelength is ~1550 nm. (c) Measured insertion loss versus peak power $P_{peak}$ of 180-fs optical pulses. (d) Excess propagation loss induced by SA of GO $\Delta SA$ versus $P_{peak}$ extracted from (c). In (a) – (d), the curves for GO-0, GO-1, and GO-2 show the results for the uncoated $Si_3N_4$ waveguides, and the hybrid waveguides with 1 and 2 layers of GO, respectively.

The linear loss was measured using continuous-wave (CW) light with a power of ~1 mW. **Figure 2a** shows the insertion loss of the fabricated devices versus wavelength. All devices exhibited nearly a flat spectral response, which suggests the absence of any material absorption or coupling loss that would generate a strong wavelength dependence. By using a cut-back method[67], we obtained a propagation loss of ~0.5 dB/cm for the $Si_3N_4$ waveguides buried in silica cladding. By comparing the $Si_3N_4$ waveguides with and without opened windows in the silica cladding, we deduced a higher propagation loss of ~3.0 dB/cm for the $Si_3N_4$ waveguides in the opened window area, which can be attributed to the mitigating effect of the silica cladding on the $Si_3N_4$ surface roughness as well as the slight roughness induced on the waveguide top surface by the cladding removal process. Finally, using these values and the measured insertion loss of the hybrid waveguides, we extracted an excess propagation loss induced by the GO films of ~3.1 dB/cm and ~6.3 dB/cm for the 1- and 2-layer devices, respectively. Such a loss induced by the GO films is about 2 orders of magnitude lower than $Si_3N_4$ waveguides integrated with graphene films[68,69], which can be attributed to the large bandgap of GO, resulting in low light absorption at near infrared wavelengths. This is a crucial advantage of GO in OPA applications where low loss is required to achieve a high net parametric gain. In principle, for GO with a bandgap > 2 eV, there should be no linear light absorption at telecom wavelengths. The linear loss of our synthesized GO films primarily originates from light absorption by localized defects, along with scattering loss due to film unevenness and imperfect interlayer contact[38,54]. To reduce the loss from these sources, we modified our GO synthesis and film coating processes by using GO solutions with high purity and optimized flake sizes.

**Figure 2b** shows the measured insertion loss versus input CW power at a wavelength of ~1550 nm. All devices showed no significant variation in insertion loss when the power was below 30 mW, indicating that the power-dependent loss induced by photo-thermal changes in the GO films was negligible within this range. This observation is consistent with our previous results where photo-thermal changes were only observed for average powers above 40 mW[56,70].

The measurement of nonlinear loss was conducted using a fiber pulsed laser (FPL) capable of generating nearly Fourier-transform limited femtosecond optical pulses centered around 1557 nm. The pulse duration and repetition rate were ~180 fs and ~60 MHz, respectively. **Figure 2c** shows the measured insertion loss versus pulse peak power $P_{peak}$. The average power of the femtosecond optical pulses was adjusted using a variable optical attenuator, ranging from 0.32 mW to 1.94 mW, which corresponds to peak powers ranging from 30 W to 180 W. The insertion loss of the hybrid waveguides decreased as the pulse peak power increased, with the 2-layer device exhibiting a more significant decrease than the 1-layer device. In contrast, the insertion loss of the uncoated $Si_3N_4$ waveguide remained constant. These results reflect that the hybrid waveguides experienced saturable absorption (SA) in the GO films, consistent with observations in waveguides incorporating graphene[68,71]. Additionally, we note that the loss changes observed were not present when using CW light with equivalent average powers. This suggests that the changes are specifically induced by optical pulses with high peak powers. In GO, the SA can be induced by the bleaching of the ground states that are associated with $sp^2$ orbitals (*e.g.*, with an energy gap of ~0.5 eV[55]) as well as the defect states. **Figure 2d** shows the SA-induced excess propagation loss ($\Delta SA$) versus pulse peak power $P_{peak}$, which was extracted from the result in **Figure 2c,** with the linear propagation loss being excluded. The negative values of $\Delta SA$ indicate that there is a decrease in loss as the peak power increases in the SA process. Such decrease in loss is beneficial for increasing the pump peak power in the OPA process, which helps improve the parametric gain.

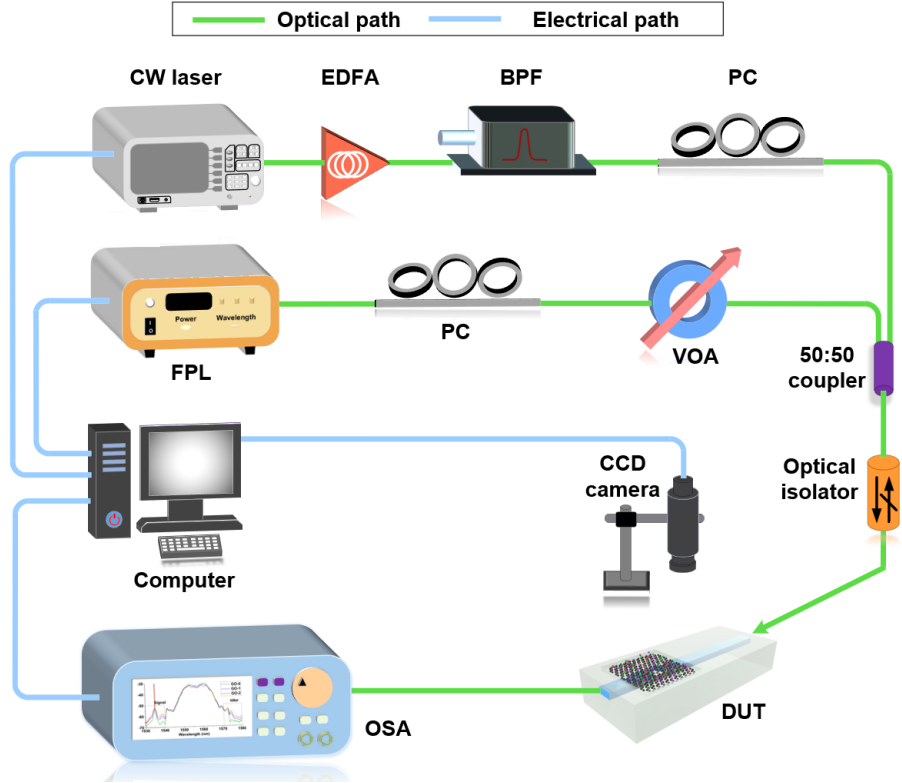

**Figure 3.** Experimental setup for OPA experiments. CW laser: continuous-wave laser. FPL: fiber pulsed laser. BPF: band-pass filter. PC: polarization controller. EDFA: erbium doped fiber amplifier. VOA: variable optical attenuator. DUT: device under test. CCD: charged-coupled device. OSA: optical spectrum analyzer.

**OPA experiments.** We conducted OPA experiments using the same devices that were fabricated and used for the loss measurements. A schematic of the experimental setup is shown in **Figure 3**. To generate the pump light required for the OPA experiments, we employed the same FPL that was used for the loss measurements. On the other hand, the signal light was generated through amplification of the CW light from a tunable laser. The pulsed pump and the CW signal were combined by a broadband 50:50 coupler and sent to the device under test (DUT) for the optical parametric process. The polarization of both signals was adjusted to TE polarized using two polarization controllers (PCs). To adjust the power of the pulsed pump, a broadband variable optical attenuator (VOA) was utilized. The output after propagation through the DUT was directed towards an optical spectrum analyzer (OSA) for analysis.

**Figure 4a** shows the optical spectra after propagation through the uncoated $Si_3N_4$ waveguide and the hybrid waveguides with 1 and 2 layers of GO. For all three devices, the input pump peak power and signal power were kept the same at $P_{peak}$ = ~180 W and $P_{signal}$ = ~6 mW, respectively. As the pump light used for the OPA experiments was pulsed, the optical parametric process occurred at a rate equivalent to the repetition rate of the FPL. As a result, both the generated idler and amplified signal also exhibited a pulsed nature with the same repetition rate as that of the FPL. The optical spectra in **Figure 4a** were analyzed to extract the parametric gain $PG$ experienced by the signal light for the three devices (see Methods). The $PG$ for the uncoated $Si_3N_4$ waveguide and the hybrid waveguides with 1 and 2 layers of GO were ~11.8 dB, ~20.4 dB, and ~24.0 dB, respectively. The hybrid waveguides exhibited higher parametric gain compared to the uncoated waveguide, and the 2-layer device had higher parametric gain than the 1-layer device. These results confirm the improved OPA performance in the $Si_3N_4$ waveguide by integrating it with 2D GO films. We also note that the hybrid devices showed greater spectral broadening of the pulsed pump caused by self-phase modulation (SPM), which is consistent with our previous observations from SPM experiments[57].

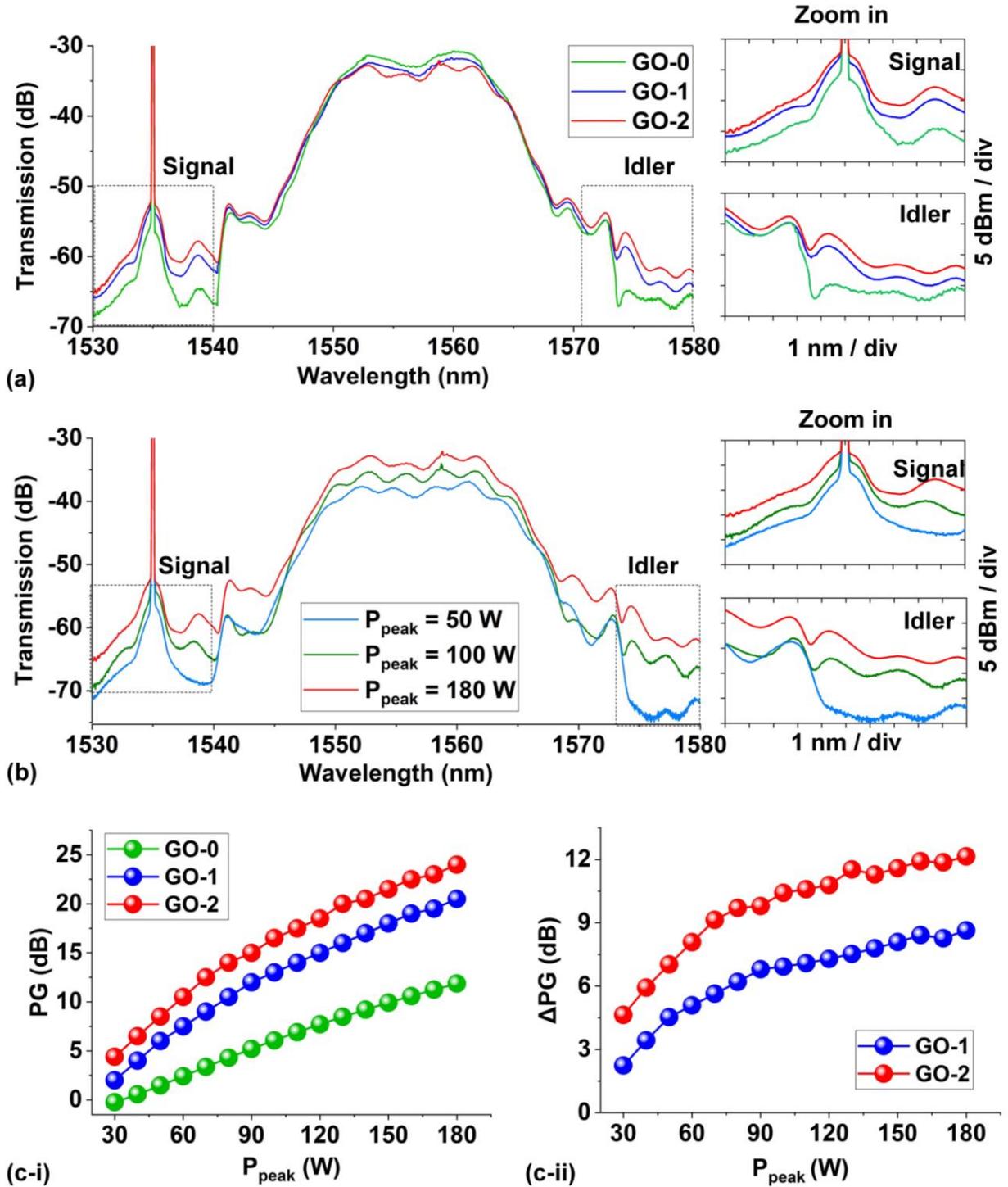

**Figure 4.** Optical parametric amplification (OPA) using a 180-fs pulsed pump and a continuous-wave (CW) signal. (a) Measured output optical spectra after propagation through uncoated (GO-0) and hybrid waveguides with 1 (GO-1) and 2 (GO-2) layers of GO. The peak power of the input pump light $P_{peak}$ was ~180 W. (b) Measured output optical spectra after propagation through the device with 2 layers of GO at different $P_{peak}$. In (a) and (b), the power of the CW signal light was $P_{signal}$ = ~6 mW, and insets show zoom-in views around the signal and idler. (c) Measured (i) parametric gain $PG$ and (ii) parametric gain improvement $\Delta PG$ versus $P_{peak}$.

The values of *PG* in **Figure 4** are the net parametric gain, over and above the waveguide loss induced by both the GO-coated and uncoated $Si_3N_4$ waveguide segments (see Methods). This is different to the "on/off" parametric gain often quoted[11,43], where the waveguide loss is excluded, resulting in higher values of parametric gain. Here, the on-off gains for the waveguides with 0, 1, and 2 layers of GO were ~13.2 dB, ~22.3 dB, and ~26.2 dB, respectively, which are only slightly higher than their corresponding net gains due to the low loss of the $Si_3N_4$ waveguides and the relatively short GO film length. Although the net gain can be increased closer to the on-off gain by reducing the waveguide loss via optimization of the fabrication processes, because the differences between the net and on-off gains are small in our case, there is not much incentive to do this. In the following, we focus our discussion on the net parametric gain *PG*. This can also ensure a fair comparison of the parametric gain improvement, as different waveguides have different waveguide loss.

**Figure 4b** shows the measured output optical spectra after propagation through the device with 2 layers of GO for different $P_{peak}$. **Figure 4c-i** shows the signal parametric gain *PG* for the uncoated and hybrid waveguides versus input pump peak power, and the parametric gain improvement Δ*PG* for the hybrid waveguides as compared to the uncoated waveguide is further extracted and shown in **Figure 4c-ii**. We varied the input pump peak power from ~30 W to ~180 W, which corresponds to the same power range used in **Figure 2d** for loss measurements. The *PG* is higher for the hybrid waveguide with 1 layer of GO compared to the uncoated waveguide, and lower than the hybrid device with 2 layers of GO. In addition, both *PG* and Δ*PG* increase with $P_{peak}$, and a maximum Δ*PG* of ~12.2 dB was achieved for the 2-layer device at $P_{peak}$ = ~180 W. Likewise, we observed similar phenomena when using lower-peak-power picosecond optical pulses for the pump, as shown in **Figure S2** of the Supplementary Information.

To evaluate the OPA performance, we conducted experiments where we varied the wavelength detuning, CW signal power, and GO film length. Except for the varied parameters, all other parameters are the same as those in **Figure 4**. In **Figure 5a**, the measured signal parametric gain *PG* and parametric gain improvement $\Delta PG$ are plotted against the wavelength detuning $\Delta\lambda$, which is defined as the difference between the CW signal wavelength $\lambda_{signal}$ and the pump center wavelength $\lambda_{pump}$. It is observed that both the *PG* and $\Delta PG$ increase as $\Delta\lambda$ changes from -12 nm to -22 nm. In **Figure 5b**, the *PG* and $\Delta PG$ are plotted against the CW signal power $P_{signal}$. As $P_{signal}$ increases, both *PG* and $\Delta PG$ remain nearly constant with only a minimal decline ($< 0.3$ dB). **Figure 5c** shows the *PG* and $\Delta PG$ versus GO film length. By measuring devices with various GO film lengths, ranging from ~0.2 mm to ~1.4 mm, we observed that those with longer GO films exhibited greater *PG* and $\Delta PG$ values. The *PG* achieved through the optical parametric process is influenced by several factors, such as the applied powers, optical nonlinearity, dispersion, and loss of the waveguides. These factors will be comprehensively analyzed in the following section.

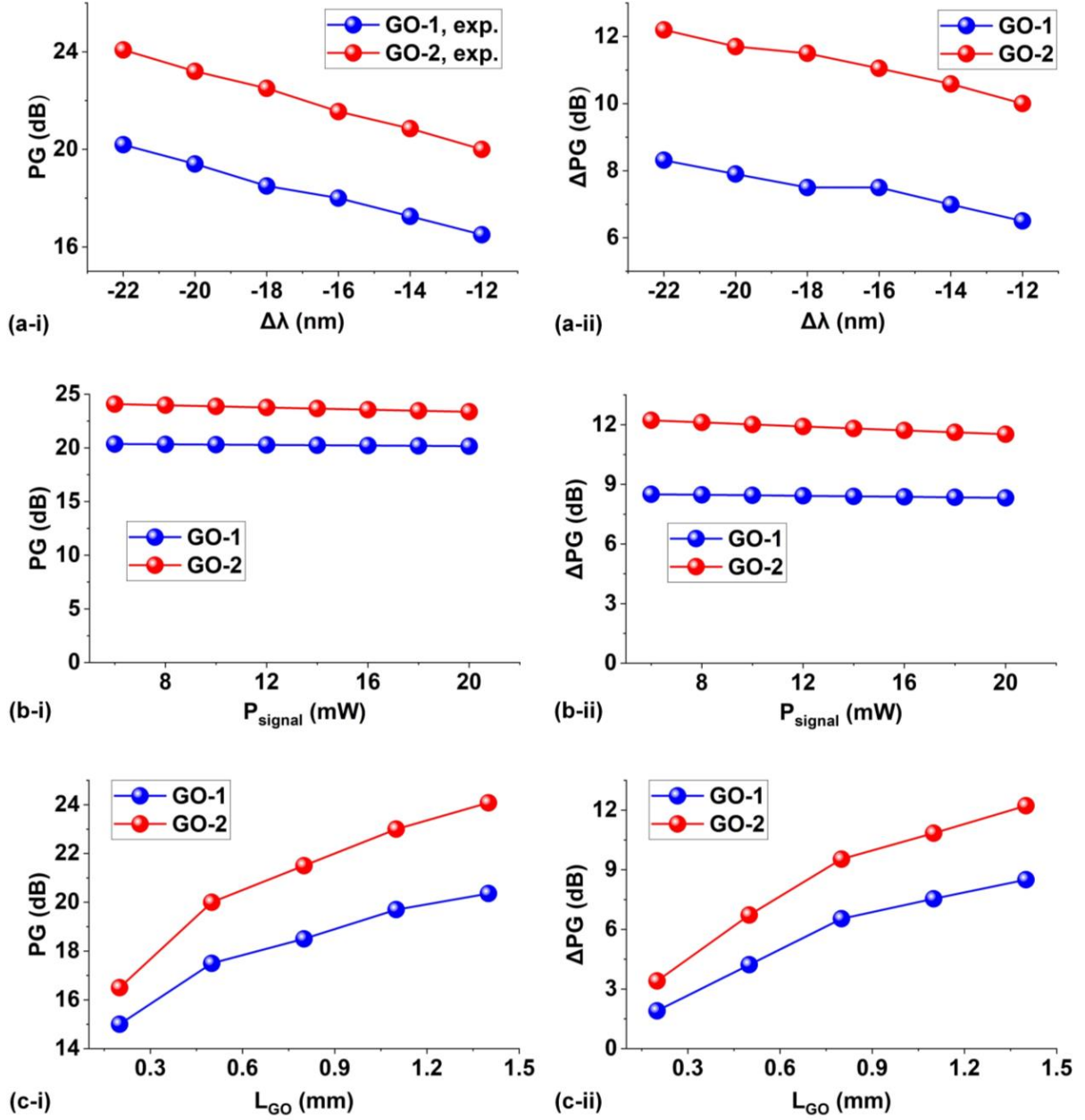

**Figure 5.** (a) Measured (i) parametric gain *PG* and (ii) parametric gain improvement $\Delta PG$ versus wavelength detuning $\Delta\lambda$. (b) Measured (i) *PG* and (ii) $\Delta PG$ versus input CW signal power $P_{signal}$. (c) Measured (i) *PG* and (ii) $\Delta PG$ versus GO film length $L_{GO}$. In (a) – (c), the peak power of the 180-fs pulsed pump centered around 1557 nm was $P_{peak}$ = ~180 W. Except for the varied parameters, all other parameters are kept the same as $\Delta\lambda$ = ~-22 nm, $P_{signal}$ = ~6 mW, and $L_{GO}$ = ~1.4 mm.

## Analysis and discussion

**Optical nonlinearity of hybrid waveguides and GO films.** We used the theory from Refs.[10,58,72] to model the OPA process in the fabricated devices (see Methods). By fitting the measured *PG* with theory, we obtained the nonlinear parameter $\gamma$ of the uncoated and hybrid waveguides. The fit $\gamma$ for the uncoated $Si_3N_4$ waveguide is ~1.11 $W^{-1}m^{-1}$, which is consistent with the previously reported values in the literature[58,73]. **Figure 6a** shows the fit $\gamma$ of the hybrid waveguides as a function of pulse peak power $P_{peak}$. For both devices with different GO film thickness, the lack of any significant variation in $\gamma$ with $P_{peak}$ indicates that the applied power has a negligible effect on the properties of the GO films. This is in contrast to the effects of light with high average optical powers, which can lead to changes in GO's properties via photo-thermal reduction[56,58]. The fit values of $\gamma$ for the devices with 1 and 2 layers of GO are ~14.5 and ~27.3 times greater than the value for the uncoated $Si_3N_4$ waveguide. These agree with our earlier work[58,59] and indicate a significant improvement in Kerr nonlinearity for the hybrid waveguides.

Based on the fit $\gamma$ for the hybrid waveguides, we further extracted the Kerr coefficient $n_2$ of the GO films (see Methods), as shown in **Figure 6b**. The extracted $n_2$ values for the films with 1 and 2 layers are similar, with the former being slightly higher than the latter. The lower $n_2$ for thicker films is likely caused by an increase in inhomogeneous defects within the GO layers and imperfect contact between multiple GO layers. The $n_2$ values for the films with 1 and 2 layers are about 5 orders of magnitude higher than that of $Si_3N_4$ (~2.62 × $10^{-19}$ $m^2$/W, obtained by fitting the result for the uncoated $Si_3N_4$ waveguide), highlighting the tremendous third-order optical nonlinearity of the GO films.

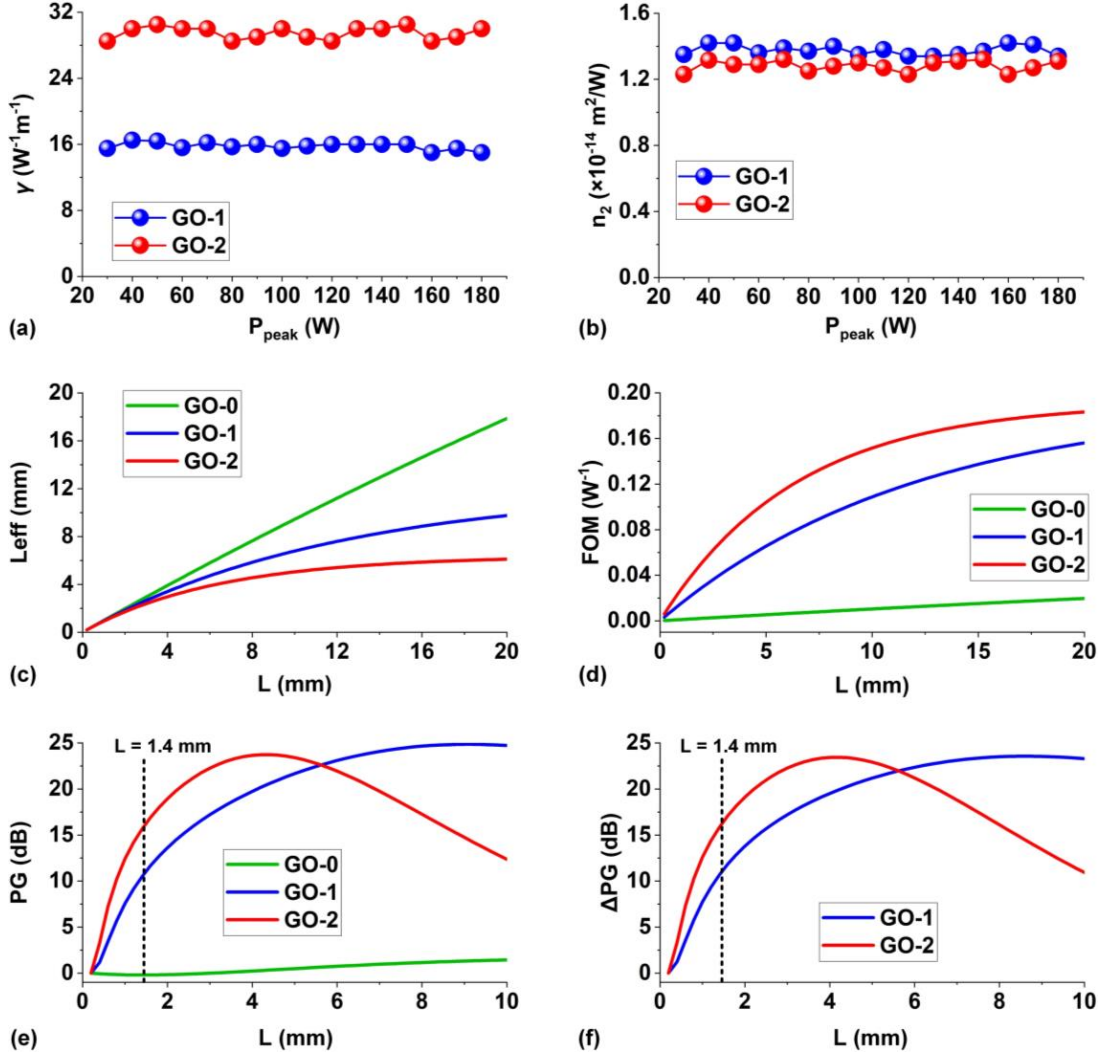

**Figure 6.** (a) Experimentally fitted nonlinear parameter $\gamma$ of hybrid waveguides with 1 (GO-1) and 2 (GO-2) layers of GO as a function of pump peak power $P_{peak}$. (b) Kerr coefficient $n_2$ of films with 1 (GO-1) and 2 (GO-2) layers of GO versus $P_{peak}$. (c) Effective interaction length $L_{eff}$ and (d) figure of merit *FOM* versus waveguide length $L$ for the uncoated (GO-0) and hybrid waveguides with 1 (GO-1) and 2 (GO-2) layers of GO. (e) Parametric gain *PG* and (f) parametric gain improvement $\Delta PG$ versus waveguide length $L$ for the uncoated $Si_3N_4$ waveguide (GO-0) and the hybrid waveguides uniformly coated with 1 (GO-1) and 2 (GO-2) layers of GO. In (e) and (f), the pump peak power, CW signal power, and the wavelength detuning are $P_{peak}$ = ~180 W, $P_{signal}$ = ~6 mW, and $\Delta\lambda$ = ~-22 nm, respectively.

We also quantitatively compare the nonlinear optical performance of the $Si_3N_4$ waveguide and the hybrid waveguides by calculating their nonlinear figure of merit *FOM*. The *FOM* is determined by balancing a waveguide's nonlinear parameter against its linear propagation loss, and can be expressed as a function of waveguide length $L$ given by:

$$FOM\ (L) = \gamma \times L_{eff}\ (L) \qquad (1)$$

where $\gamma$ is the waveguide nonlinear parameter and $L_{eff}(L) = [1 - exp(-\alpha \times L)]/\alpha$ is the effective interaction length, with $\alpha$ denoting the linear loss attenuation coefficient. Note that the nonlinear figure of merit defined in **Eq. (1)**, which gives the maximum nonlinear effect induced per unit of Watt, allows for comparison of the nonlinear optical performance of optical waveguides made from different materials. This is distinct from the nonlinear figure of merit commonly used for comparing the nonlinear optical performance of a single material, which is defined as $n_2/(\lambda \cdot \beta_{TPA})$[36], with $n_2$, $\lambda$, and $\beta_{TPA}$ denoting the Kerr coefficient, wavelength, TPA coefficient, respectively.

**Figure 6c** shows $L_{eff}$ versus $L$ for the $Si_3N_4$ waveguide and the hybrid waveguides with 1 and 2 layers of GO. The $Si_3N_4$ waveguide has a higher $L_{eff}$ due to its comparably lower linear propagation loss. **Figure 6d** shows the *FOM* versus $L$ for the three waveguides. Despite having a lower $L_{eff}$, the hybrid waveguides exhibit a higher *FOM* than the $Si_3N_4$ waveguide, owing to the significantly improved nonlinear parameter $\gamma$ for the hybrid waveguides. This indicates that the impact of enhancing the optical nonlinearity is much greater than the degradation caused by the increase in loss, resulting in a significant improvement in the device's overall nonlinear optical performance.

For the hybrid waveguides that we measured in the OPA experiments, only a specific section of the waveguides was coated with GO films. In **Figures 6e** and **6f**, we compare *PG* and *ΔPG* versus waveguide length $L$ for the hybrid waveguides uniformly coated with GO films, respectively, which were calculated based on the fit $\gamma$ values (at $P_{peak}$ = ~180 W) in **Figure 6a**. The pump peak power, CW signal power, and wavelength detuning were $P_{peak}$ = 180 W, $P_{signal}$ = 6 mW, and $\Delta\lambda$ = -22 nm, respectively – the same as those in **Figure 4a**. The corresponding results for the uncoated $Si_3N_4$ waveguide are also shown for comparison. The 2-layer device has higher *PG* and *ΔPG* values for $L <$ ~5.7 mm but lower values for $L >$ ~5.7 mm, reflecting the trade-off between the increase in optical nonlinearity and waveguide loss. At $L$ = 1.4 mm, the 1-layer and 2-layer devices achieve

*PG* of ~10.5 dB and ~15.6 dB, respectively. When compared to waveguides that have patterned GO films of the same length as those used in our OPA experiments, their total *PG* (including those provided by both the ~1.4-mm-long GO-coated section and the ~18.6-mm-long uncoated section) are ~20.4 dB and ~24.0 dB, respectively. This highlights the dominant role of the GO-coated section in providing the parametric gain, as well as the fact that a further improvement in $\Delta PG$ could be obtained by increasing the length of the GO-coated segments.

**Performance improvement by optimizing parameters.** Based on the OPA modeling (see Methods) and the fit parameters in **Figure 6**, we further investigate the margin for performance improvement by optimizing the parameters.

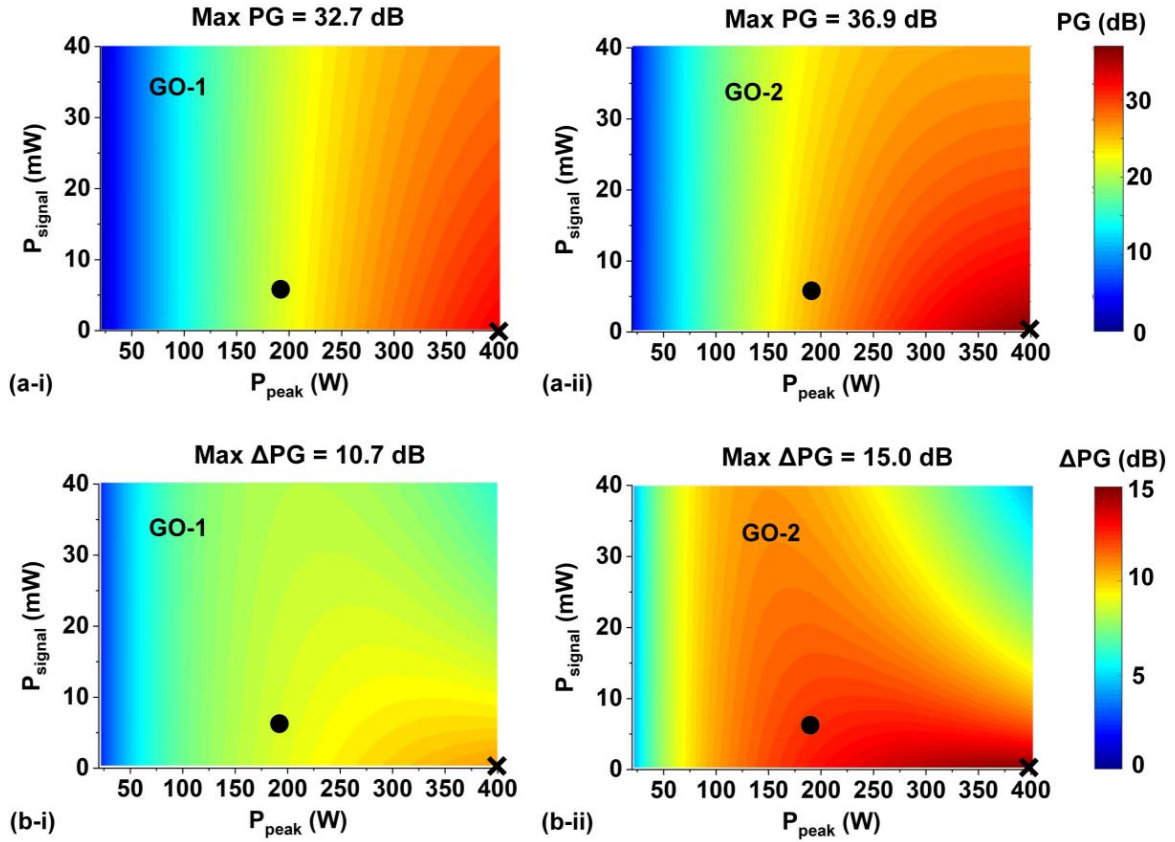

**Figure 7.** (a) Simulated parametric gain *PG* versus input pump peak power $P_{peak}$ and CW signal power $P_{signal}$. (b) Simulated parametric gain improvement $\Delta PG$ versus $P_{peak}$ and $P_{signal}$. In (a) and (b), (i) and (ii) show the results for the hybrid waveguides with 1 and 2 layers of GO (GO-1, GO-2), respectively. The black points mark the OPA experimental results, and the black crossing mark the results corresponding to the maximum values of *PG* and $\Delta PG$. The wavelength detuning and the GO film length are $\Delta\lambda$ = -22 nm and $L_{GO}$ = 1.4 mm, respectively.

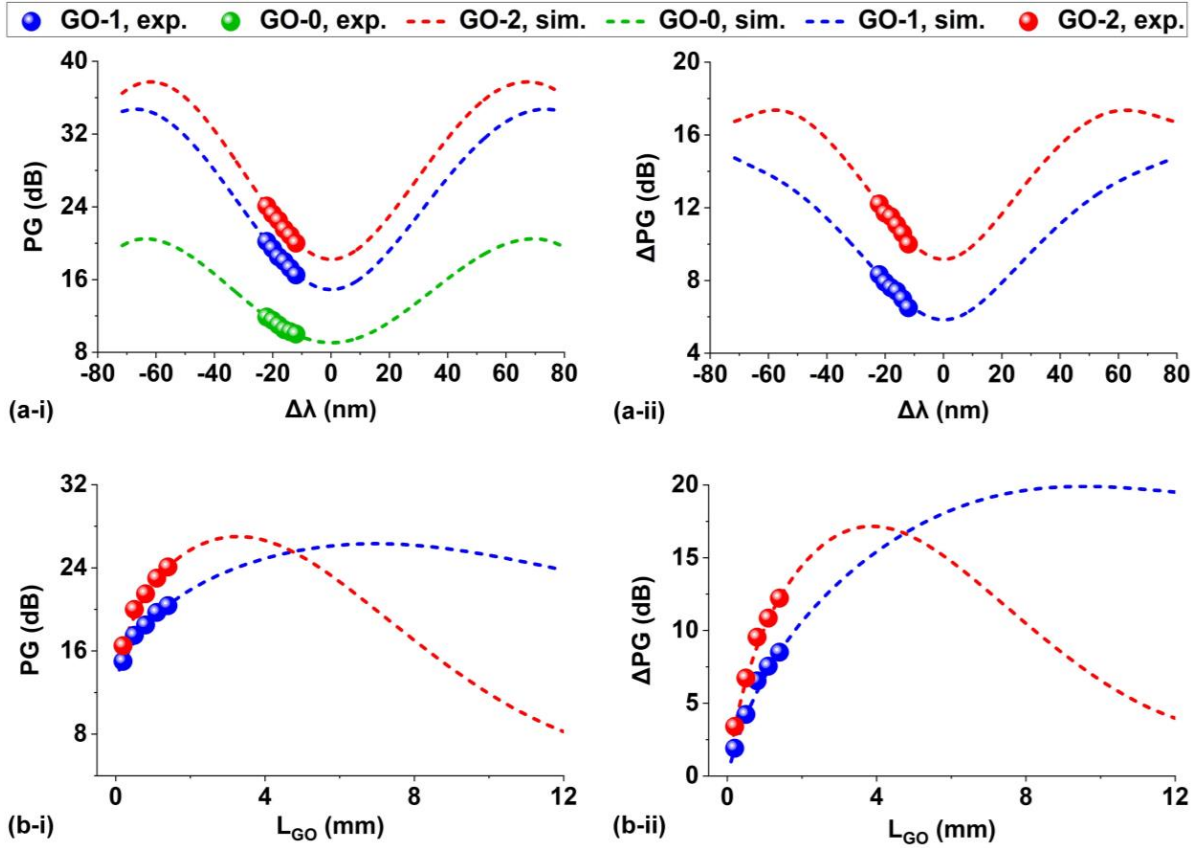

**Figure 8.** (a) Simulated (i) parametric gain *PG* and (ii) parametric gain improvement Δ*PG* versus wavelength detuning Δλ. (b) Simulated (i) *PG* and (ii) Δ*PG* versus GO coating length $L_{GO}$. In (a) and (b), the measured and fit results are shown by the data points and the dashed curves, respectively. The pump peak power and the signal power are $P_{peak}$ = 180 W and $P_{signal}$ = 6 mW, respectively. In (a), $L_{GO}$ = 1.4 mm. In (b), Δλ = -22 nm.

**Figure 7a** shows the calculated *PG* for the hybrid waveguides versus pulse peak power $P_{peak}$ and CW signal power $P_{signal}$. The corresponding results for Δ*PG* are shown in **Figure 7b**. In each figure, (i) and (ii) show the results for the devices with 1 and 2 layers of GO, respectively. The black points mark the experimental results in **Figure 4**, and the black crossings mark the results corresponding to the maximum values of *PG* or Δ*PG*. As can be seen, both *PG* and Δ*PG* increase with $P_{peak}$ but slightly decrease with $P_{signal}$, showing agreement with the trends observed in the experimental results. For the device with 1 layer of GO, the maximum *PG* of ~32.7 dB and Δ*PG* of ~10.7 dB are achieved at $P_{peak}$ = 400 W and $P_{signal}$ = 1 mW. Whereas for the device with 2 layers of GO, the maximum *PG* and Δ*PG* are ~36.9 dB and ~15.0 dB at the same $P_{peak}$ and $P_{signal}$, respectively. This reflects that there is a large room for improvement by further optimizing the pulse peak power and the CW signal power. In our experiments, the maximum output power of our FPL limited the applied pulse peak power. In addition, we opted to avoid using excessively low CW signal power due to two reasons. First, the CW signal power does not exert a significant influence on PG. Second, as the power of the output pulsed signal diminishes with the decrease of the input CW signal power, it becomes increasingly challenging to extract PG accurately.

**Figure 8a** shows the calculated $PG$ and $\Delta PG$ versus wavelength detuning $\Delta\lambda$. The dashed curves were calculated based on the fit result at $\Delta\lambda = -22$ nm, and the data points mark the measured results in **Figure 5a**. The curves with an 'M' shape are consistent with the results in Refs.[3,10,11], reflecting the anomalous dispersion of these waveguides. The experimental data points match closely with the simulation curves, thereby confirming the consistency between our experimental results and theory. For the device with 1 layer of GO, the maximum $PG$ of ~34.7 dB and $\Delta PG$ of ~14.7 dB are achieved at $\Delta\lambda = $ ~-67.0 nm and ~-80.0 nm, respectively. Whereas for the 2-layer device, the maximum $PG$ of ~37.6 dB and $\Delta PG$ of ~17.3 dB are achieved at $\Delta\lambda = $ ~-61.8 nm and -~57.8 nm, respectively. These results highlight the significant potential for improvement through further optimization of the wavelength detuning. In our experiments, the range of wavelength detuning was limited by the operation bandwidth of the erbium-doped fiber amplifier used to amplify the CW signal power.

We also investigate the performance improvement by optimizing the GO film length $L_{GO}$. **Figure 8b** shows the calculated $PG$ and $\Delta PG$ versus $L_{GO}$. The dashed curves were calculated based on the fit result at $L_{GO} = 1.4$ mm, and the data points mark the measured results in **Figure 5a**. For the device with 1 layer of GO, the maximum $PG$ of ~26.3 dB and $\Delta PG$ of ~19.9 dB are achieved at $L_{GO} = $ ~7 mm and ~9.7 mm, respectively. Whereas for the device with 2 layers of GO, the maximum $PG$ of ~27.0 dB and $\Delta PG$ of ~17.2 dB are achieved at $L_{GO} = $ ~3.3 mm and ~3.9 mm, respectively. These results suggest that the OPA performance can be improved by further optimizing the length of the GO film. In our experiments, the lengths of the GO films were restricted by the size of the opened windows on the silica cladding (as shown in **Figure 1d**). Aside from optimizing the GO film length, we would anticipate even higher values of *PG* and Δ*PG* for devices with an increased number of GO layers at $L_{GO}$ = 1.4 mm, similar to what we observed in our previous nonlinear optics experiments[56,57]. This is due to the considerably increased optical nonlinearity of devices with thicker GO films. However, such an increase in optical nonlinearity is accompanied by a rise in loss, making it imperative to balance the trade-off between them.

We investigate the performance by optimizing both Δ$\lambda$ and $L_{GO}$ simultaneously (see **Figure S4** of the Supplementary Information), finding that the 1-layer device has a maximum *PG* of ~37.4 dB and maximum Δ*PG* of ~31.5 dB, while the 2-layer device reaches *PG* up to ~37.8 dB and Δ*PG* up to ~27.3 dB. In addition, by further increasing the pump peak power from 180 W to 400 W, even higher performance is achieved, with the 1-layer device reaching a maximum *PG* of ~43.7 dB and maximum Δ*PG* of ~40.1 dB, and the 2-layer device a maximum *PG* of ~43.8 dB and maximum Δ*PG* of ~37.3 dB (see **Figure S5** of the Supplementary Information). According to these simulation results, it is found that if both Δ$\lambda$ and $L_{GO}$ are optimized simultaneously, there is not much difference between the maximum *PG* for the 1- and 2- layer devices. However, the 1-layer device still yields a slightly higher Δ*PG* because of its lower loss compared with the 2-layer device. For this reason, devices coated with more GO layers will have lower maximum Δ*PG*.

We investigate the improvement in *PG* and Δ*PG* by optimizing the coating position of the GO films (see **Figure S6** of the Supplementary Information), as well as the influence of the SA of GO on the OPA performance (see **Figure S7** of the Supplementary Information). We find that although optimizing the coating position can lead to further improvements in *PG* and Δ*PG*, the extent of these improvements is not as substantial as those achieved through optimization of Δ$\lambda$ and $L_{GO}$. In addition, we find that the SA of GO has a positive impact on enhancing *PG* and Δ*PG*, especially for devices with thicker GO films.

While this work is the first report of parametric gain in waveguides integrated with 2D materials, we compare our results for the OPA performance of Si$_3$N$_4$-GO waveguides with theoretical calculated OPA performance of Si$_3$N$_4$-graphene waveguides. By using the graphene material parameters from Refs.[51,74,75], we performed theoretical simulations similar to those in **Figures 7** and **8**. The results show that the hybrid waveguide with monolayer graphene can achieve a maximum Δ*PG* of ~3.6 dB at $P_{peak}$ = 180 W and $P_{signal}$ = 6 mW. This is much lower than the comparable value for the hybrid waveguide with monolayer GO (*i.e.*, ~12.2 dB), highlighting the excellent performance of GO-based OPA devices. The main reason for this is due to the very high linear loss of the graphene film.

Note that we have demonstrated enhanced nonlinear optics in three CMOS-compatible integrated nonlinear optical platforms – silicon, Si$_3$N$_4$, and high index doped silica[54]. This work represents the first study of parametric gain in any of these platforms integrated with GO films. We chose Si$_3$N$_4$ waveguides because it has a large bandgap of ~5.0 eV[54] that yields low TPA in the near infrared region. Also, there is a strong need for enhancing the optical nonlinearity of Si$_3$N$_4$ waveguides because its intrinsic $n_2$ is quite low at only about 10 times of silica. In silicon, on the other hand, the needs are quite different since the $n_2$ is already quite high. Here the need is more for decreasing the nonlinear absorption. For high index doped silica the challenge lies in the fact that because the mode sizes tend to be fairly large, the overlap with the GO films is much smaller, limiting the enhancement in the optical nonlinearity. We expect that the improved OPA performance achieved through integration of 2D GO films will extend beyond the Si$_3$N$_4$ platform to all the three platforms and this will form the basis of future studies.

In this work, we demonstrate enhanced OPA performance in GO-Si$_3$N$_4$ hybrid waveguides by using optical pulses with high peak powers. Further optimization of the GO film properties, *e.g.*, reducing the loss and increasing the optical nonlinearity, could potentially enable meaningful applications such as optical microcomb generation, where high parametric gain needs to be achieved for a CW light with a low peak power[24,36]. In that case, rather than pumping with high power pulses, the increase of pump power would be accomplished through intracavity resonant enhancement of the CW pump. This will be an important subject of our future work. In principle, GO films with a bandgap > 2 eV should have negligible linear light absorption at near-infrared wavelengths. The loss of practically synthesized GO films primarily results from light absorption due to localized defects, as well as scattering loss arising from uneven film surfaces and imperfect layer contact. By optimizing the film synthesis and coating processes, there is still significant room to reduce the loss arising from these sources. Another potential limitation arises from the reduction of GO when subjected to CW lights with high average powers. Previously, we discovered that employing an electrochemical method to adjust the oxidation degree of GO allowed it to maintain a high Kerr nonlinearity while significantly enhancing its stability at high laser powers[76]. This offers a potential solution to mitigate this concern.

## Conclusion

In summary, we experimentally demonstrate significantly improved OPA performance in Si$_3$N$_4$ waveguides integrated with 2D GO films compared to uncoated waveguides. We fabricate GO-Si$_3$N$_4$ hybrid waveguides with precise control of the thickness, length, and position of the GO films. Detailed OPA measurements are performed for the fabricated devices using a pulsed pump and CW signal. The results show that up to ~24.0 dB parametric gain is achieved for the hybrid devices, representing a ~12.2-dB improvement relative to the device without GO. Based on the experimental results, the influence of the pump / signal power, wavelength detuning, and GO film thickness /

length on the OPA performance is theoretically analyzed, showing that further improvement can be achieved by optimizing these parameters. We calculate that a parametric gain of ~37.8 dB and a parametric gain improvement of ~31.5 dB should be possible by optimizing the wavelength detuning and GO film length, and even higher to ~43.8 dB by increasing the pump peak power to 400 W. Our study provides valuable insights into the promising potential of on-chip integration of 2D GO films for enhancing the OPA performance of photonic integrated devices, of benefit to many nonlinear optical applications.

## Materials and methods

**Fabrication of $Si_3N_4$ waveguides.** The $Si_3N_4$ waveguides were fabricated via CMOS compatible processes[77,78]. First, a $Si_3N_4$ film was deposited on a silicon wafer with a 3-μm-thick wet oxidation layer on its top surface, using a low-pressure chemical vapor deposition (LPCVD) method. The deposition was carried out in two steps involving a twist-and-grow process, resulting in a crack-free film. Next, waveguides were created using 248-nm deep ultraviolet lithography followed by fluorocarbon-based dry etching with $CF_4$/$CHF_3$/Ar, which resulted in a low sidewall surface roughness for the waveguides. After waveguide patterning, we employed a multi-step, chemical-physical, in-situ annealing sequence using $H_2$, $O_2$, and $N_2$ to further reduce the loss of the $Si_3N_4$ waveguides. Subsequently, a silica upper cladding was deposited to encapsule the $Si_3N_4$ waveguides via multi-step low-temperature oxide deposition at 400 °C. This was achieved through a low-rate deposition of a liner, followed by the filling of the silica layer using high-density plasma enhanced chemical vapor deposition (HD-PECVD). Finally, we employed lithography and dry etching to create windows on the silica cladding extending to the top surface of the $Si_3N_4$ waveguides.

**Synthesis and coating of GO films.** Before GO film coating, a GO solution with small GO flake size (< 100 nm) was prepared through chemical oxidation of graphite using a modified Hummers method[63]. After oxidation, the hydrophobic graphite was converted to hydrophilic GO by attaching OCFGs, which allows the GO flakes to dissolve in water. After intensive sonication using a Branson Digital Sonifier, the GO flakes in solution fragmented into small negatively charged nanoflakes with monolayer thickness, which remained well-dispersed to prevent aggregation.

The coating of 2D layered GO films was achieved by using a transfer-free method that allows for layer-by-layer GO film deposition with precise control of the film thickness, as we did previously[79,80]. During the coating process, four steps for in-situ assembly of monolayer GO films were repeated to construct multi-layered films on the fabricated $Si_3N_4$ chips with opened windows, including (i) immerse substrate into a 2.0% (w/v) aqueous PDDA (Sigma-Aldrich) solution; (ii) rinse with a stream of deionized distilled water and dry with $N_2$; (iii) immerse the PDDA-coated substrate into GO solution; and (iv) rinse with a stream of deionized water and dry with $N_2$. After the film coating, the chip was dried in a drying oven.

The electrostatic force enables conformal film coating onto complex structures such as wire waveguides and gratings with a high degree of uniformity. However, achieving conformal GO film coating in narrow slot regions with widths <100 nm and heights >200 nm remains a challenge. This limitation is primarily attributed to the size of the GO nanoflakes employed for self-assembly, which was typically ~50 nm in our prepared GO solution. By modifying the GO synthesis methods and using more vigorous ultrasonication, GO flakes with smaller sizes can be obtained, which can potentially mitigate this issue.

The solution-based, layer-by-layer GO film coating method demonstrates excellent scalability and compatibility with different integrated optical platforms such as silicon, silicon nitride, and high index doped silica. Unlike techniques involving film transfer, where the coated areas are constrained by the lateral size of the exfoliated 2D films, the coverage area achievable through our self-assembly method is constrained only by the substrate and the solution container dimensions[12].

This makes it particularly proficient at coating large areas. By using plasma oxidation, the removal of GO films that have been coated onto integrated devices can be effortlessly accomplished. This allows for the reclamation of integrated chips and the subsequent reapplication of new GO films. Apart from using opened windows to control the length and positioning of GO films on integrated waveguides, this can also be achieved by using lithography followed by lift-off[38,56].

**Extracting parametric gain from the measured optical spectra.** We used the same methods as those in Refs.[11,43] to extract the signal parametric gain from the measured optical spectra we obtained through OPA experiments. The peak power of the pulsed signal after propagation through the fabricated devices was derived from the measured output optical spectra according to:

$$P_{signal,\ peak} = \frac{\iint P_{signal,\ out}(\lambda) d\lambda}{f_{rep} \times T} \tag{2}$$

where $P_{signal,\ out}(\lambda)$ is the average output power spectrum of the signal as a function of wavelength $\lambda$, $f_{rep}$ is the repetition rate of the FPL, and $T$ is the pulse width. In our calculation of $P_{signal,\ peak}$, the power residing in the CW signal line was subtracted from the spectrum of $P_{signal,\ out}(\lambda)$.

After deriving $P_{signal,\ peak}$, the signal parametric gain $PG$ was calculated as:

$$PG\ (\text{dB}) = 10 \times \log_{10}(P_{signal,\ peak} / P_{signal}) \tag{3}$$

where $P_{signal}$ is the CW signal power at the input of the waveguide. According to **Eq. (3)**, the $PG$ in our discussion is the net gain over and above the waveguide loss (including that induced by both the $Si_3N_4$ waveguide and the GO film). In contrast, the on/off parametric gain is defined as[11,43]

$$PG_{on\text{-}off}\ (\text{dB}) = 10 \times \log_{10}(P_{signal,\ peak} / P_{signal,\ out}) \tag{4}$$

where $P_{signal,out}$ is the CW signal power at the output of the waveguide when the pump is turned off. The parametric gain calculated using **Eq. (4)** is higher than that calculated using **Eq. (3)** since $P_{signal,out}$ is lower than $P_{signal}$.

**OPA Modeling.** The third-order optical parametric process in the GO-coated $Si_3N_4$ waveguides was modeled based on the theory from Refs.[10,58,72]. Assuming negligible depletion of the pump and signal powers due to the generation of the idler, and considering only the short wavelength idler, the coupled differential equations for the dominant degenerate FWM process can be given by[10,65]

$$\frac{dA_p(z)}{dz} = -\frac{\alpha_p}{2}A_p(z) + j\gamma_p \left[|A_p(z)|^2 + 2|A_s(z)|^2 + 2|A_i(z)|^2\right]A_p(z)$$

$$+ j2\gamma_p A_p^*(z)A_s(z)A_i(z)exp(j\Delta\beta z) \tag{5}$$

$$\frac{dA_s(z)}{dz} = -\frac{\alpha_s}{2}A_s(z) + j\gamma_s \left[|A_s(z)|^2 + 2|A_p(z)|^2 + 2|A_i(z)|^2\right]A_s(z)$$

$$+ j\gamma_s A_i^*(z)A_p^2(z)exp(-j\Delta\beta z) \tag{6}$$

$$\frac{dA_i(z)}{dz} = -\frac{\alpha_i}{2}A_i(z) + j\gamma_i \left[|A_i(z)|^2 + 2|A_p(z)|^2 + 2|A_s(z)|^2\right]A_i(z)$$

$$+ j\gamma_i A_s^*(z)A_p^2(z)exp(-j\Delta\beta z) \tag{7}$$

where $A_{p,s,i}$ are the amplitudes of the pump, signal and idler waves along the z axis, which is defined as the light propagation direction, $\alpha_{p,s,i}$ are the loss factor including both the linear loss and the SA-induced nonlinear loss, $\Delta\beta = \beta_s + \beta_i - 2\beta_p$ is the linear phase mismatch, with $\beta_{p,s,i}$ denoting the propagation constants of the pump, signal and idler waves, and $\gamma_{p,s,i}$ are the waveguide nonlinear parameters. In our case, where the wavelength detuning range was small ($\leq$ 10 nm), the linear loss and the nonlinear parameter are assumed to be constant, *i.e.*, $\alpha_p = \alpha_s = \alpha_i = \alpha$, $\gamma_p = \gamma_s = \gamma_i = \gamma$.

In **Eqs. (5) − (7)**, the dispersions $\beta_{p,s,i}$ were calculated via commercial mode solving software using the refractive index *n* of layered GO films measured by spectral ellipsometry. Given that the photo-thermal changes are sensitive to the average power in the hybrid waveguides, which was below 2 mW for the femtosecond optical pulses studied here, they were considered negligible. By numerically solving **Eqs. (5) – (7)**, the *PG* was calculated via

$$PG \text{ (dB)} = 10 \times \log_{10}[|A_s(L)|^2/|A_s(0)|^2] \quad (8)$$

where $L$ is the length of the $Si_3N_4$ waveguide (*i.e.*, 20 mm). For our devices with patterned GO films, the waveguides were divided into uncoated $Si_3N_4$ (without GO films) and hybrid (with GO films) segments with different $\alpha$, $\gamma$ and $\beta_{p,s,i}$. The differential equations were solved for each segment, with the output from the previous segment as the input for the subsequent segment.

**Extracting $n_2$ of GO films.** The Kerr coefficient $n_2$ of the layered GO films is extracted from the nonlinear parameter $\gamma$ of the hybrid waveguides according to:[56,72]

$$\gamma = \frac{2\pi}{\lambda_c} \frac{\iint_D n_0^2(x,y) n_2(x,y) S_z^2 \, dxdy}{\left[\iint_D n_0(x,y) S_z \, dxdy\right]^2} \quad (9)$$

where $\lambda_c$ is the pulse central wavelength, $D$ is the integral of the optical fields over the material regions, $S_z$ is the time-averaged Poynting vector calculated using Lumerical FDTD commercial mode solving software, $n_0(x, y)$ and $n_2(x, y)$ are the linear refractive index and $n_2$ profiles over the waveguide cross section, respectively. The values of $n_2$ for silica and $Si_3N_4$ used in our calculation were $2.60 \times 10^{-20}$ m$^2$/W[36] and $2.62 \times 10^{-19}$ m$^2$/W, respectively. The latter was obtained by fitting the experimental results for the uncoated $Si_3N_4$ waveguide.

**Acknowledgements**

This work was supported by the Australian Research Council Centre of Excellence Project in Optical Microcombs for Breakthrough Science (No. CE230100006), the Australian Research Council Discovery Projects Programs (DP190103186, FT210100806), Linkage Program (LP210200345), the Swinburne ECR-SUPRA program, the Industrial Transformation Training Centers scheme (Grant No. IC180100005), the Beijing Natural Science Foundation (No. Z180007), the Agence Nationale de la Recherche (ANR) (Grant No. MIRSiCOMB, ANR-17-CE24-0028), and the H2020 European Research Council (ERC) (Grant No. GRAPHICS, 648546). The work of Christian Grillet and Christelle Monat was supported by the International Associated Laboratory in Photonics between France and Australia (LIA ALPhFA).

**Conflict of interest**

The authors declare no competing financial interest.

**Supplementary information**

Figures S1−S7 and references